\newcommand*{\ATLASLATEXPATH}{}
 
\documentclass[PAPER, texlive=2016, UKenglish, txfonts, cernpreprint]{\ATLASLATEXPATH atlasdoc}
 
\usepackage{\ATLASLATEXPATH atlaspackage}
\usepackage{\ATLASLATEXPATH atlasbiblatex}

\usepackage{\ATLASLATEXPATH atlasphysics}
\usepackage{\ATLASLATEXPATH atlasheavyion}
\usepackage{\ATLASLATEXPATH atlasprocess}
\usepackage{\ATLASLATEXPATH atlasjetetmiss}
\graphicspath{{logos/}{figures/}}

\usepackage{verbatim}

\newcommand{\dcaT}{\mbox{$d_{0}$}\xspace}

\newcommand{\rhodef}{\mbox{$(p^{\textrm{ID}} - p^{\textrm{MS}})/p^{\textrm{ID}}$}\xspace}

\newcommand*{\sumETFcal}{\ensuremath{\Sigma E_{\text{T}}^{\ce{FCal}}}\xspace}

\newcommand{\dphiPsi}[1]{\mbox{$#1|\phi-\Psi_{#1}|$}\xspace}

\AtlasTitle{Measurement of azimuthal anisotropy of muons from charm and bottom hadrons in Pb+Pb collisions at $\sqrt{s_\mathrm{NN}} = 5.02$~TeV with the ATLAS detector }

\AtlasAbstract{
Azimuthal anisotropies of muons from charm and bottom hadron decays are measured in Pb+Pb collisions at $\sqrt{s_\mathrm{NN}}= 5.02~\mathrm{TeV}$.   The data were collected with the ATLAS detector at the Large Hadron Collider in 2015 and 2018 with integrated luminosities of $0.5~\mathrm{nb}^{-1}$ and $1.4~\mathrm{nb^{-1}}$, respectively.
The kinematic selection for heavy-flavor muons requires transverse momentum $4 < p_\mathrm{T} < 30~\mathrm{GeV}$ and pseudorapidity $|\eta|<2.0$. The dominant sources of
muons in this $p_\mathrm{T}$ range are semi-leptonic decays of charm and bottom hadrons.
These heavy-flavor muons are separated from light-hadron decay muons and punch-through hadrons using the momentum imbalance between the measurements in the tracking detector and in the muon spectrometers.
Azimuthal anisotropies, quantified by flow coefficients, are measured via the event-plane method for inclusive heavy-flavor muons as a function of the muon $p_\mathrm{T}$ and in intervals of Pb+Pb collision centrality.
Heavy-flavor muons are separated into contributions from charm and bottom hadron decays
using the muon transverse impact parameter with respect
to the event primary vertex.   Non-zero elliptic ($v_{2}$) and triangular ($v_{3}$) flow coefficients are extracted for charm and bottom muons, with the charm muon coefficients larger than those for bottom muons for all Pb+Pb collision centralities.   The results indicate substantial modification to the charm and bottom quark angular distributions through interactions in the quark-gluon plasma produced in these Pb+Pb collisions, with smaller modifications for the bottom quarks as expected theoretically due to their larger mass.
}
 
\author{The ATLAS Collaboration}
 
\AtlasRefCode{HION-2019-11}
 
\AtlasJournalRef{Phys. Lett. B 807 (2020) 135595}
\AtlasDOI{10.1016/j.physletb.2020.135595}

\PreprintIdNumber{CERN-EP-2019-274}
\hypersetup{pdftitle={ATLAS document},pdfauthor={The ATLAS Collaboration}}

\begin{document}
 
\maketitle

\section{Introduction}

The paradigm for the time evolution of heavy-ion collisions at the Relativistic Heavy Ion Collider (RHIC) and the Large Hadron Collider (LHC) involves the formation and hydrodynamic expansion of a region of hot and dense quark--gluon plasma (QGP) with a small ratio of the shear viscosity to entropy density.  In this paradigm, the QGP is considered to be a nearly perfect fluid ~\cite{Romatschke:2017ejr,Heinz:2013th}.
Initial geometric inhomogeneities of the QGP are translated into momentum anisotropies of the final-state hadrons via large pressure gradients.
Extensive measurements of light-hadron azimuthal anisotropies have been performed, in which the single-particle azimuthal distributions are expressed in terms of a Fourier expansion:
 
\begin{equation}
\frac{\mathrm{d}N}{\mathrm{d}\phi} \propto 1 + 2 \sum_{n=1}^{\infty} v_{n} \cos(n(\phi-\Psi_{n})),
\label{eqn:fourier}
\end{equation}
 
where the event-plane angle, $\Psi_n$, specifies the orientation of the initial density profile in the transverse plane~\cite{Voloshin:1994mz},
and Fourier coefficients, $v_n$, quantify the magnitude of the modulation with respect to the event-plane angle.
The second- and third-order $v_{n}$ coefficients are referred to as elliptic ($v_2$) and triangular ($v_3$) flow coefficients, respectively, with the term `flow' invoking the hydrodynamic paradigm.

Heavy-flavour (charm and bottom) quarks have masses much larger than the temperature of the QGP ($m_{c,b} > T)$, with maximum temperatures at early times
ranging between 300 and $500~\MeV$~\cite{Habich:2014jna}.
Thus, thermal production of heavy quarks during the QGP phase is highly suppressed.   Instead, heavy quarks are typically produced at the earliest times via high-momentum-transfer collisions between incoming partons.
Once created, the heavy quarks persist throughout the dynamical time evolution of the QGP and thus act as sensitive probes of the hot and dense medium.
 
Owing to their larger masses, radiative energy loss of heavy quarks in the QGP is suppressed relative to that of light quarks~\cite{Dokshitzer:2001zm}.
However, it was still postulated that charm quarks interact strongly enough to flow with the QGP~\cite{Batsouli:2002qf}.
Experimental data at RHIC and then at the LHC reveals that heavy-quark hadrons, as well as their decay leptons, have transverse momentum (\pt) distributions that are strongly modified by the QGP relative to observations in proton--proton (\pp) collisions~\cite{HION-2015-06,CMS-HIN-16-001,CMS-HIN-16-016,CMS-HIN-16-011,Acharya:2018hre,Adam:2018inb}.
Charm hadrons~\cite{Acharya:2017qps,CMS-HIN-16-007} and heavy-quark hadron decay leptons~\cite{HION-2015-06,Adam:2016ssk} are also observed to have significant azimuthal anisotropies, suggesting that they participate in the overall collective flow of the medium.
For recent reviews of heavy-flavour measurements in heavy-ion collisions, see Refs.~\cite{Averbeck:2013oga,Dong:2019byy,Andronic:2015wma}.
 
For $\pt\lesssim \textrm{4--6}~\GeV$, it was proposed that heavy quarks can be described via a Langevin approach with drag and diffusion terms~\cite{Moore:2004tg}.   Modified \pt distributions and azimuthal anisotropies of $D$ mesons have been used to constrain heavy-quark transport coefficients~\cite{Cao:2018ews,Rapp:2018qla}.   Other models of heavy-flavor kinematics in the QGP, including a Boltzmann approach, have also been explored~\cite{PhysRevLett.110.112301,Alberico:2011zy,PhysRevC.84.024908,PhysRevC.79.044906,Zhang:2005ni}.  At higher momenta $\pt \gtrsim 5-10~\GeV$, heavy-quark energy loss is thought to dominate, with collisional and induced radiative processes both contributing
~\cite{Wicks:2005gt}.   At intermediate \pT hadronization effects can be important as azimuthal anisotropies for the deconfined heavy-quark is transferred to the heavy-flavor hadron~\cite{He:2014cla}.
There are numerous theoretical predictions for the azimuthal anisotropies of bottom quarks, e.g.\ in Refs.~\cite{Katz:2019fkc,Zigic:2018ovr,Ke:2018tsh}; however, only limited experimental data are currently available.   Precision experimental data for \pt distributions and azimuthal anisotropies is crucial as this over-constrains the calculations that depend on the heavy-quark to QGP coupling as well as the QGP space-time evolution.
 
The flow coefficients $v_{2}$, $v_{3}$, and $v_{4}$ of inclusive heavy-flavour muon production, which includes both muons from charm hadron decays (``charm muons'') and muons from bottom hadron decays (``bottom muons''), have been measured by the ATLAS experiment~\cite{HION-2015-06} and ALICE experiment~\cite{Adam:2015pga} in \PbPb collisions at $\sqn=2.76~\TeV$.
The measurement indicates significant elliptic flow for heavy-flavour muons with $4 < \pt < 10~\GeV$.
Recently, the heavy-flavour muon $v_{2}$ has also been measured in
high-multiplicity $\sqrt{s} = 13~\TeV$ \pp collisions~\cite{HION-2018-23}.   Unlike the earlier Pb+Pb measurement, the \pp measurement examined charm muons and bottom muons separately, finding a non-zero $v_2$ for charm muons while the $v_2$ for bottom muons is consistent with zero within uncertainties.
 
In the measurement presented in this paper,
the procedure of the
previous $\sqn=2.76~\TeV$ \PbPb analysis using the event-plane method is followed~\cite{HION-2015-06}, and is extended to extract separate flow coefficients for charm and bottom muons. These measurements extends the previously published ones to the higher $\sqn=5.02~\TeV$ \PbPb beam energy, using a larger event sample provided by the 2015 and 2018 combined data sets.    The larger data sample enables measurements over a larger
momentum range $4 < \pt < 30~\GeV$ for inclusive heavy-flavour muons.   It also allows the separation of the inclusive heavy-flavour muons into charm and bottom contributions.  Results for charm and bottom muon elliptic $v_2$ and triangular $v_3$ flow coefficients are presented as a function of muon \pt for various ranges of overlap between the colliding
nuclei, referred to as ``centrality''.
 
\section{ATLAS detector}

\newcommand{\AtlasCoordFootnote}{
ATLAS uses a right-handed coordinate system with its origin at the nominal interaction point (IP)
in the centre of the detector and the \(z\)-axis along the beam pipe.
The \(x\)-axis points from the IP to the centre of the LHC ring,
and the \(y\)-axis points upwards.
Cylindrical coordinates \((r,\phi)\) are used in the transverse plane,
\(\phi\) being the azimuthal angle around the \(z\)-axis.
The pseudorapidity is defined in terms of the polar angle \(\theta\) as \(\eta = -\ln \tan(\theta/2)\).
Angular distance is measured in units of \(\Delta R \equiv \sqrt{(\Delta\eta)^{2} + (\Delta\phi)^{2}}\).}
 
The ATLAS detector~\cite{PERF-2007-01,ATLAS-TDR-2010-19,PIX-2018-001} at the LHC is a multipurpose particle detector
with a forward--backward symmetric cylindrical geometry and a near \(4\pi\) coverage in
solid angle.\footnote{\AtlasCoordFootnote}
It consists of an inner tracking detector surrounded by a thin superconducting solenoid
providing a \SI{2}{\tesla} axial magnetic field, electromagnetic and hadron calorimeters, and a muon spectrometer.
The inner tracking detector (ID) covers the pseudorapidity range \(|\eta| < 2.5\).
It consists of silicon pixel, silicon microstrip, and transition radiation tracking (TRT) detectors.
The calorimeter system covers the pseudorapidity range \(|\eta| < 4.9\).
Within the region \(|\eta|< 3.2\), electromagnetic calorimetry is provided by barrel and
endcap high-granularity lead/liquid-argon (LAr) calorimeters,
with an additional thin LAr presampler covering \(|\eta| < 1.8\),
to correct for energy loss in material upstream of the calorimeters.
Hadronic calorimetry is provided by the steel/scintillator-tile calorimeter,
segmented into three barrel structures within \(|\eta| < 1.7\), and two copper/LAr hadronic endcap calorimeters.
The solid angle coverage is completed with forward copper/LAr and tungsten/LAr calorimeter modules (FCal)
optimised for electromagnetic and hadronic measurements respectively.
The muon spectrometer (MS) comprises separate trigger and
high-precision tracking chambers, covering $|\eta| < 2.4$ and $|\eta| < 2.7$ respectively,
and measures the deflection of muons in a magnetic field generated by superconducting air-core toroids.
The field integral of the toroids ranges between \num{2.0} and \SI{6.0}{\tesla\metre}
across most of the detector. The minimum-bias trigger scintillators (MBTS) detect charged
particles over $2.07<|\eta|<3.86$ using two hodoscopes of 12 counters positioned at $z=\pm$3.6~m.
The zero-degree calorimeters (ZDC) measure neutral particles at pseudorapidities $|\eta| \geq 8.3$
and consist of layers of alternating quartz rods and tungsten plates.
A two-level trigger system~\cite{TRIG-2016-01} is used to select events.
The first-level trigger (L1) is implemented in hardware and uses a subset of the detector information
to reduce the accepted rate to below \SI{100}{\kHz}.
This is followed by a software-based high-level trigger (HLT) stage that
reduces the accepted event rate to 1--\SI{4}{\kilo\hertz}
depending on the data-taking conditions during \PbPb collisions.

\section{Event selection}
 
Data used in this analysis were recorded with the ATLAS detector in 2015 and 2018 from \PbPb collisions at $\sqn=5.02~\TeV$ with integrated luminosities of $0.5~\mathrm{nb}^{-1}$ and  $1.4~\mathrm{nb}^{-1}$, respectively.
Events were selected online using a set of muon triggers that require a muon at the HLT stage with \pT larger than 3, 4, 6, or 8~\GeV
~\cite{TRIG-2016-01}.
The muon trigger selecting \pt $> ~8~\GeV$ sampled the full luminosity in both 2015 and 2018,
while triggers with lower \pt thresholds were prescaled to reduce the overall data rate.  Thus the higher-threshold triggers are utilised at a given muon \pt to sample a larger fraction of the full luminosity.
The resulting sampled luminosities are $0.3~\mathrm{nb}^{-1}$, $0.6~\mathrm{nb}^{-1}$, and $1.9~\mathrm{nb}^{-1}$ for muons with $4 < \pt < 6~\GeV$, $6 < \pt < 8~\GeV$ and $\pt > 8~\GeV$, respectively.
The selected events are further required to satisfy offline minimum-bias \PbPb collision criteria to reject pile-up based on a combination of the total transverse energy measured in the FCal, denoted by \sumETFcal, and the ZDC energy.
 
The centrality of each \PbPb event is characterised by its \sumETFcal.  For the results shown here, the minimum-bias \sumETFcal distribution is divided into percentiles ordered from the most central (large \sumETFcal, small impact parameter) to the most peripheral (small \sumETFcal, large impact parameter):  0--10\%, 10--20\%, 20--30\%, 30--40\%, and 40--60\%, where 0--100\% corresponds to the total \PbPb inelastic cross section.    A Monte Carlo Glauber~\cite{Miller:2007ri} calculation
is used to characterise each centrality interval~\cite{HION-2016-06}.   The above centrality intervals have an average number of participating nucleons $\avgNpart = 358.8 \pm 2.3$, $264.1 \pm 2.9$, $189.2 \pm 2.8$, $131.4 \pm 2.6$, and $70.5 \pm 2.2$ respectively.

Muons with $4 < \pt < 30$~\GeV\ and $|\eta|<2.0$ reconstructed in both the ID and the MS are selected and required to pass `medium' selection requirements, detailed in Ref.~\cite{PERF-2015-10}, without any requirement on the number of TRT hits, due to the high occupancy in heavy-ion running.
Candidate muons are required to be matched with a muon reconstructed by the event trigger.
Each muon is assigned a weight given by the inverse of the reconstruction and trigger efficiency for the specific muon kinematics.

The muon reconstruction and trigger efficiencies are determined using the \Jmm tag-and-probe method as detailed in Ref.~\cite{PERF-2015-10}.
The muon reconstruction efficiency is factorized as the product of ID and MS reconstruction efficiencies.
The ID reconstruction efficiency is obtained from \PbPb data directly,
while the MS reconstruction efficiency is obtained both from \PbPb data and by overlaying \PbPb minimum-bias events on simulated $\Jpsi$ produced by
\PYTHIAV{8}~\cite{Sjostrand:2014zea} with the CTEQ6L1~\cite{Pumplin:2002vw} parton distribution functions, using the same tag-and-probe method.
The events in overlay simulations are then given weights such that the \sumETFcal distribution matches the muon-triggered \PbPb data distribution.
The MS reconstruction efficiency obtained from simulation is used as the central value for MS reconstruction, with additional data-to-MC scale factors applied to account for residual differences between data and overlay simulation.
The same \Jmm tag-and-probe method is used to determine the muon trigger efficiency.
The central value of the single-muon trigger efficiency correction is obtained from simulations without overlaying \PbPb minimum-bias events.
Additional correction factors of order 1--10\% are obtained from data and applied to the selection to correct for trigger detector performance differences between data and simulation, as well as the centrality dependence of the muon trigger efficiency in \PbPb data.
 
\section{Analysis procedure}
 
The analysis follows the event-plane method for measuring flow coefficients as used in
previous ATLAS measurements~\cite{HION-2015-06,HION-2016-06} and is briefly summarised here.   Each \PbPb event has a geometric orientation of the impact parameter vector, and the event can also have a tilt relative to that due to fluctuations in the geometry of the resulting QGP.
In any particular \PbPb collision, one can estimate the orientation, represented by the FCal-determined $n^\mathrm{th}$-order event-plane angle $\Psi_n$.
The azimuthal distribution of transverse energy deposited in the forward and backward rapidity FCal is used to determine the event plane.
A comparison of event planes, as measured separately in the forward-rapidity FCal and the backward-rapidity FCal, enables a determination of the event-plane resolution $\mathrm{Res}\{n\Psi_{n}\}$ as detailed in Ref.~\cite{HION-2016-06}.
In each \PbPb centrality interval and in each muon \pt selection, the muons are divided into a finite number of intervals in $\phi - \Psi_{n}$, where $\phi$ is the azimuthal angle of the muon.
As different harmonic orders are orthogonal to each other, the Fourier decomposition of the angular distribution (introduced in \Eqn({\ref{eqn:fourier}})) at a given order $n$ can be expressed as
 
\begin{equation*}
\frac{1}{N_{X}^\mu}\frac{\mathrm{d}N_{X}^\mu}{\mathrm{d}(n(\phi-\Psi_{n}))} = 1+2 v_{n}^{\mathrm{raw}} \cos \big( n(\phi-\Psi_{n}) \big),
\end{equation*}
 
where the $v_{n}^{\mathrm{raw}}$ are the raw flow coefficients and
the $N_{X}^\mu$ are the extracted yields for the muons of interest. Three types of signal are considered in this measurement
($X$ = charm, bottom, and inclusive heavy-flavour).
The final $v_n$ coefficients are obtained by correcting for the event-plane resolution:
$v_n = {v_n^\mathrm{raw}} / {\mathrm{Res}\{n\Psi_{n}\}}$.
The leading sources of background contribution in the selected muon samples are muons from decay-in-flight and punch-through of $\pi$ and $K$  ($\pi/ K$ background) and muons from non-heavy-flavour components such as direct quarkonia, low-mass resonances, $\tau$-leptons and $W/Z$ decays (labeled ``light/onia background'').
Other sources of background from hadronic showers and fake muons are found to be very small and are only considered in systematic uncertainties.
 
Similarly to previous ATLAS publications~\cite{STDM-2011-07,HION-2015-06,HION-2018-23}, different sources of muons are separated using two variables.  The first is the momentum imbalance, $\rho = \rhodef$, where $p^\mathrm{ID}$ is the muon momentum measured in the ID, and $p^\mathrm{MS}$ is that measured in the MS corrected for the energy loss inside the calorimeter.
Real muons have a $\rho$ distribution peaked around zero while the $\pi/ K$ background has a broader $\rho$ distribution that is shifted toward higher values.
The different shapes of the $\rho$ distribution for the $\pi/ K$ background and other muons enable the isolation of the $\pi/ K$ background muons.
The analysis is repeated using the transverse momentum imbalance, as opposed to the total momentum imbalance $\rho$, and no difference is observed.
The second variable is the transverse impact parameter, $d_0$, relative to the event's primary vertex~\cite{ATL-PHYS-PUB-2015-026}.
Charm and bottom muons have different \dcaT distributions due to the different decay lengths of charm and bottom hadrons.
 
A two-step fit in $\rho$ and \dcaT is performed in data, using $\rho$ and \dcaT line-shape templates for different sources of muons obtained from simulations.
First, the yields of inclusive heavy-flavour and $\pi/ K$ background muons are extracted from a fit to the $\rho$ distribution.
The relative yields of light/onia background muons and inclusive heavy-flavour muons are fixed to the fractions obtained from \PYTHIAV{8} simulations, which are approximately $4\%$ on average.
Then, with the extracted $\pi/ K$ background yields fixed, a fit to the \dcaT distribution is performed to determine the relative fraction of charm and bottom muons within the yield of inclusive heavy-flavour muons.
 
The muon ID momentum resolution in \PYTHIAV{8} simulations
overlaid with minimum-bias \PbPb events
is found to be worse than the resolution in \PbPb data.   Thus, the $\rho$ templates are obtained from simulation without \PbPb event overlay.
The single-muon ID and MS momentum responses in the \PYTHIAV{8} simulation are shifted and smeared in order to match those in \PbPb data.
The single-muon momentum shift and smearing parameters are determined by comparing the invariant mass resolution of simulated \Jmm events in \PYTHIAV{8} with that from \PbPb data at different centralities.
The charm and bottom muon $\rho$ templates are determined from
multijet hard-scattering \pp collision events at $\sqrt{s} =13~\TeV$  filtered on the presence of a generator-level muon
in \PYTHIAV{8} with parameter values as in the A14 tune~\cite{ATL-PHYS-PUB-2014-021} and using the NNPDF23LO parton distribution functions~\cite{Ball:2012cx}.
The $\pi/K$ background $\rho$ templates are obtained from non-diffractive QCD simulations of \pp collisions at $\sqrt{s} =13~\TeV$ in \PYTHIAV{8}, also with the A14 tune and NNPDF23LO parton distribution functions.
The $\rho$ templates for the light/onia background contribution are obtained from simulations of direct $\Jpsi$ in \pp collisions at $\sqrt{s} =5.02~\TeV$.
No differences in the template shapes were observed between simulations at $\sqrt{s} = 13~\TeV$ and $\sqrt{s} = 5.02~\TeV$.
The signal muon $\rho$ distribution shape shows no obvious dependence on muon \pt, but is found to be broader in the endcap muon detector and in most central events due to poorer muon momentum resolution in the ID.
 
As the \dcaT resolution is sensitive to the primary vertex position resolution, the \dcaT templates are all obtained from \PYTHIAV{8} simulations at $\sqrt{s} =5.02~\TeV$ overlaid with minimum-bias \PbPb events to best approximate the primary vertex resolution.
The distributions of \dcaT are shifted and smeared to remove residual differences between overlay simulations and \PbPb data.
The \dcaT shift and smearing parameters are found by comparing the distributions of high-quality prompt-tracks between \PbPb data and overlay simulations.
The charm and bottom muon \dcaT templates are obtained from
multijet hard-scattering simulations filtered on the presence of a generator-level muon, whereas the $\pi/K$ background \dcaT templates are from non-diffractive QCD simulations, and the templates for light/onia background are obtained from direct $\Jpsi$ simulations.
The signal muon \dcaT distribution shape shows no dependence on muon \pt or event centrality but a moderate dependence on parent charm and bottom hadron \pt due to the strong correlation between decay length and particle velocity.
Additional reweighting is applied to the charm and bottom muon signal samples to match the input charm and bottom hadron \pt spectra to those measured in \PbPb collisions by ALICE~\cite{Acharya:2018hre} and CMS~\cite{CMS-HIN-16-001,CMS-HIN-16-025}.

\begin{figure}
\centering
\includegraphics[width=0.4\linewidth]{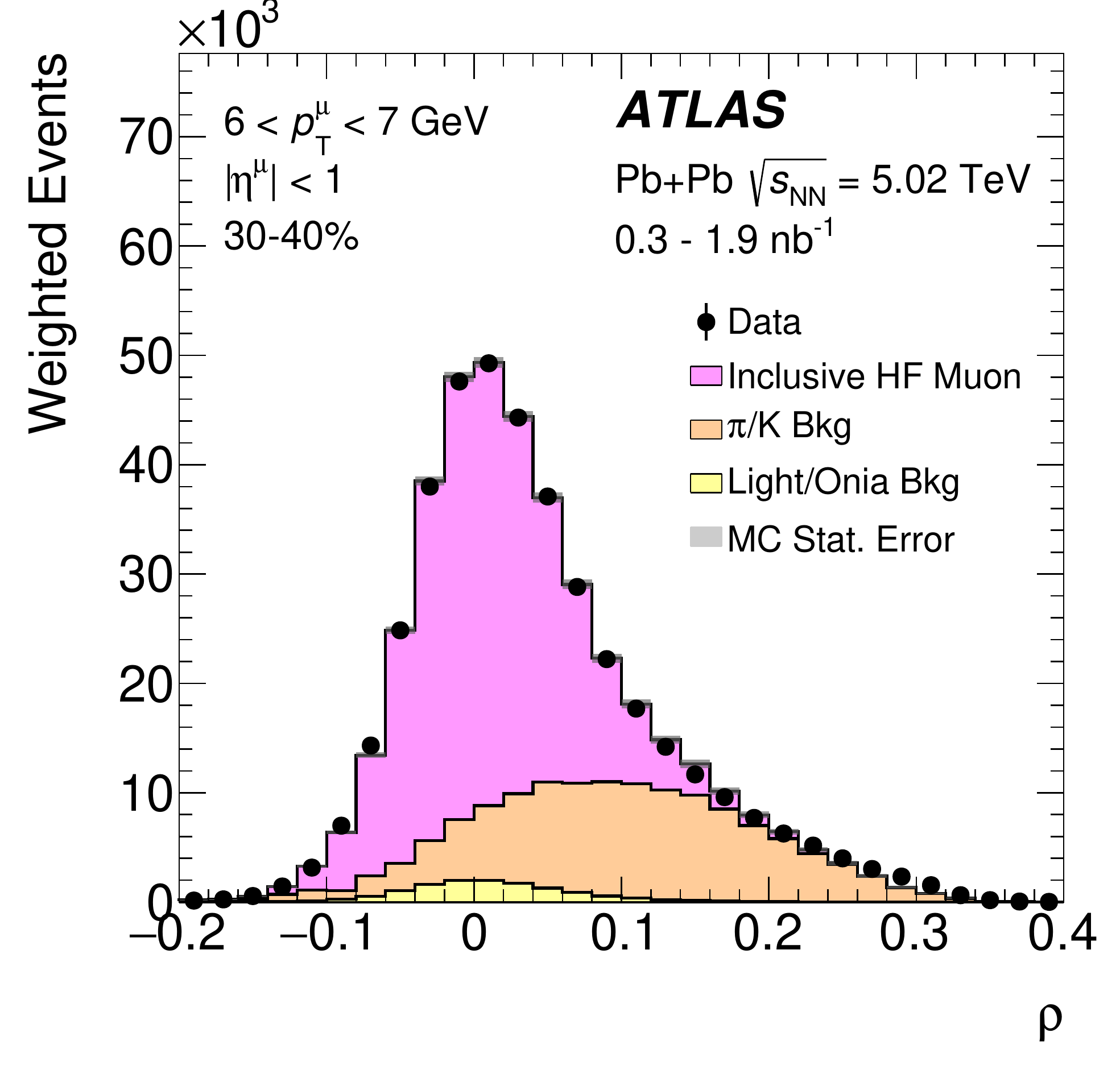}
\includegraphics[width=0.4\linewidth]{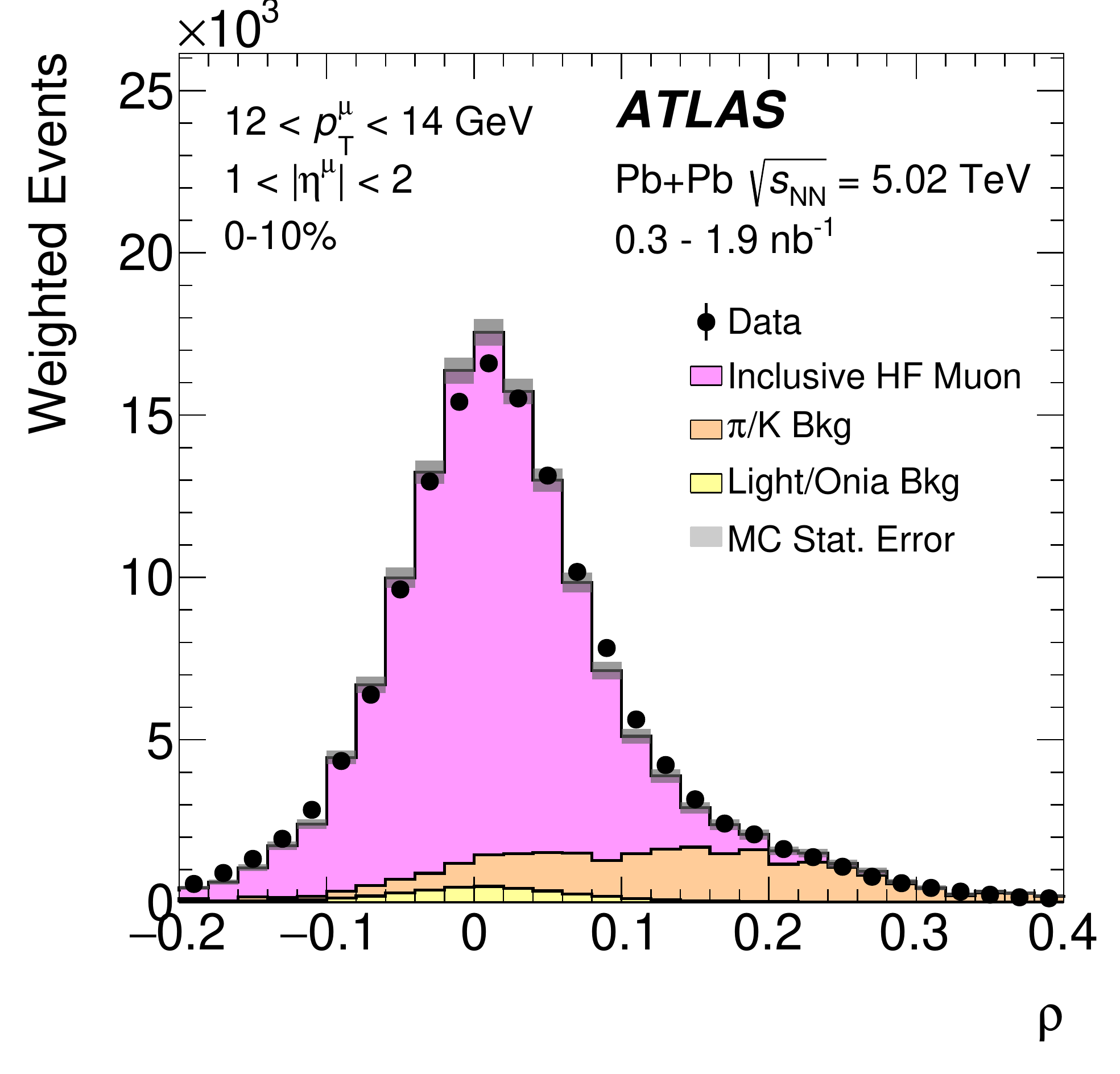}
\includegraphics[width=0.4\linewidth]{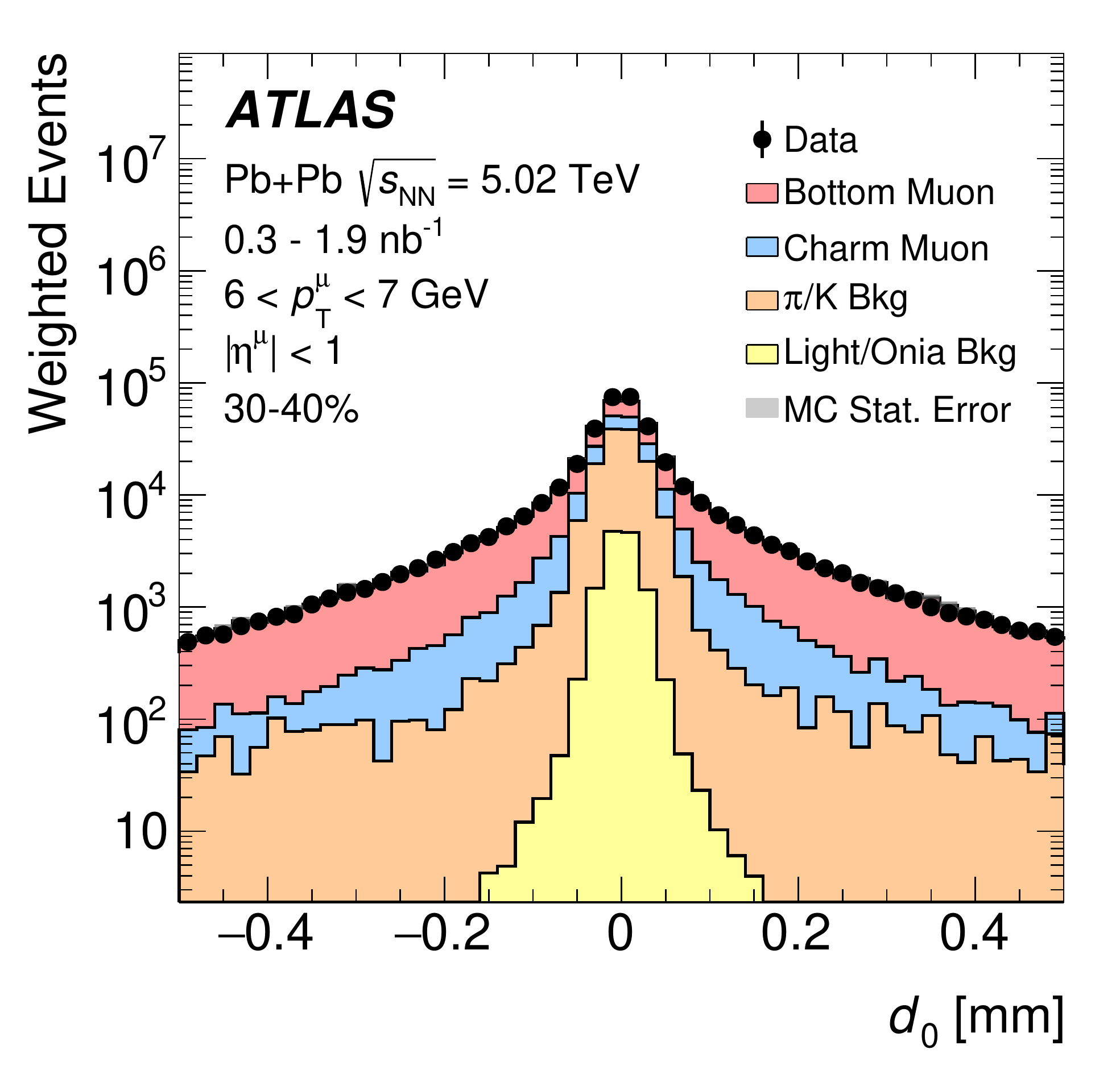}
\includegraphics[width=0.4\linewidth]{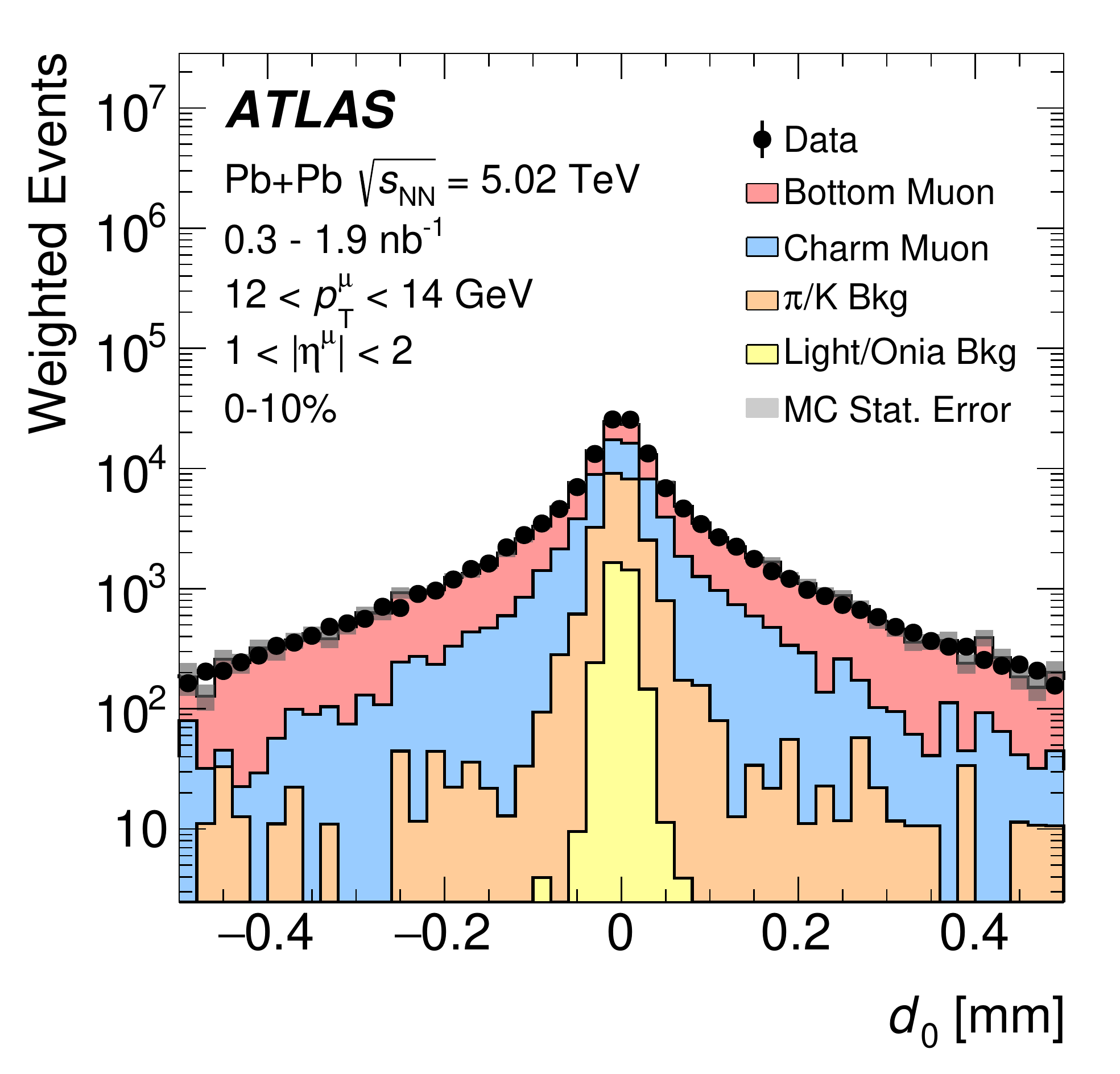}
\caption{Example fits to $\rho$ (top) and \dcaT (bottom) for muons with $6 < \pt < 7~\GeV$ and $|\eta| < 1$ in 30--40\% centrality \PbPb collisions (left) and $12 < \pt < 14~\GeV$ and $1 < |\eta| < 2$ in 0--10\% centrality \PbPb collisions (right) both integrated over $\dphiPsi{n}$. Muon trigger and reconstruction efficiency corrections are applied to the data.
}
\label{fig:example_fit}
\end{figure}

The fits are performed independently in differential intervals of muon \pt, centrality, $\dphiPsi{n}$ and two intervals of muon $\eta$.
The two muon $\eta$ intervals ($|\eta| < 1$ and $1 < |\eta| < 2$) are fitted independently to minimize residual data/MC difference in the barrel and endcap muon detectors separately, and then combined to obtain charm, bottom, and inclusive heavy-flavour muon yields in the given \pt, centrality, and $\dphiPsi{n}$ intervals as reported in the results.
Fluctuations in the simulation templates are incorporated in the fitting procedure.
Examples of selected fits in $\rho$ and $\dcaT$ based on simulation templates are shown in \Fig{\ref{fig:example_fit}} for two different muon \pt selections ($6 < \pt < 7~\GeV$ and $12 < \pt < 14~\GeV$) integrated over $\dphiPsi{n}$.

\begin{figure}[htb]
\centering
\includegraphics[width=0.40\linewidth]{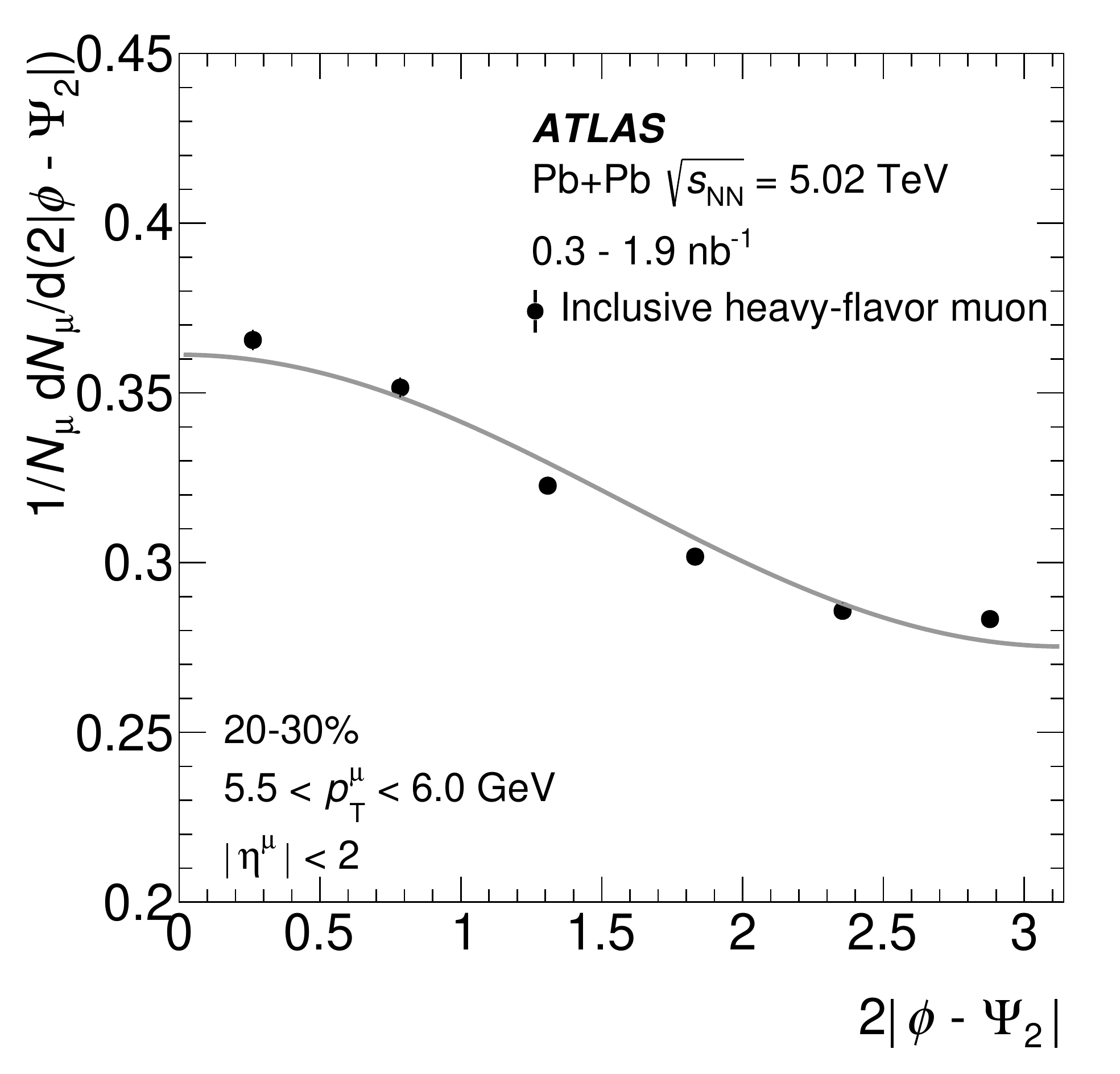}
\includegraphics[width=0.40\linewidth]{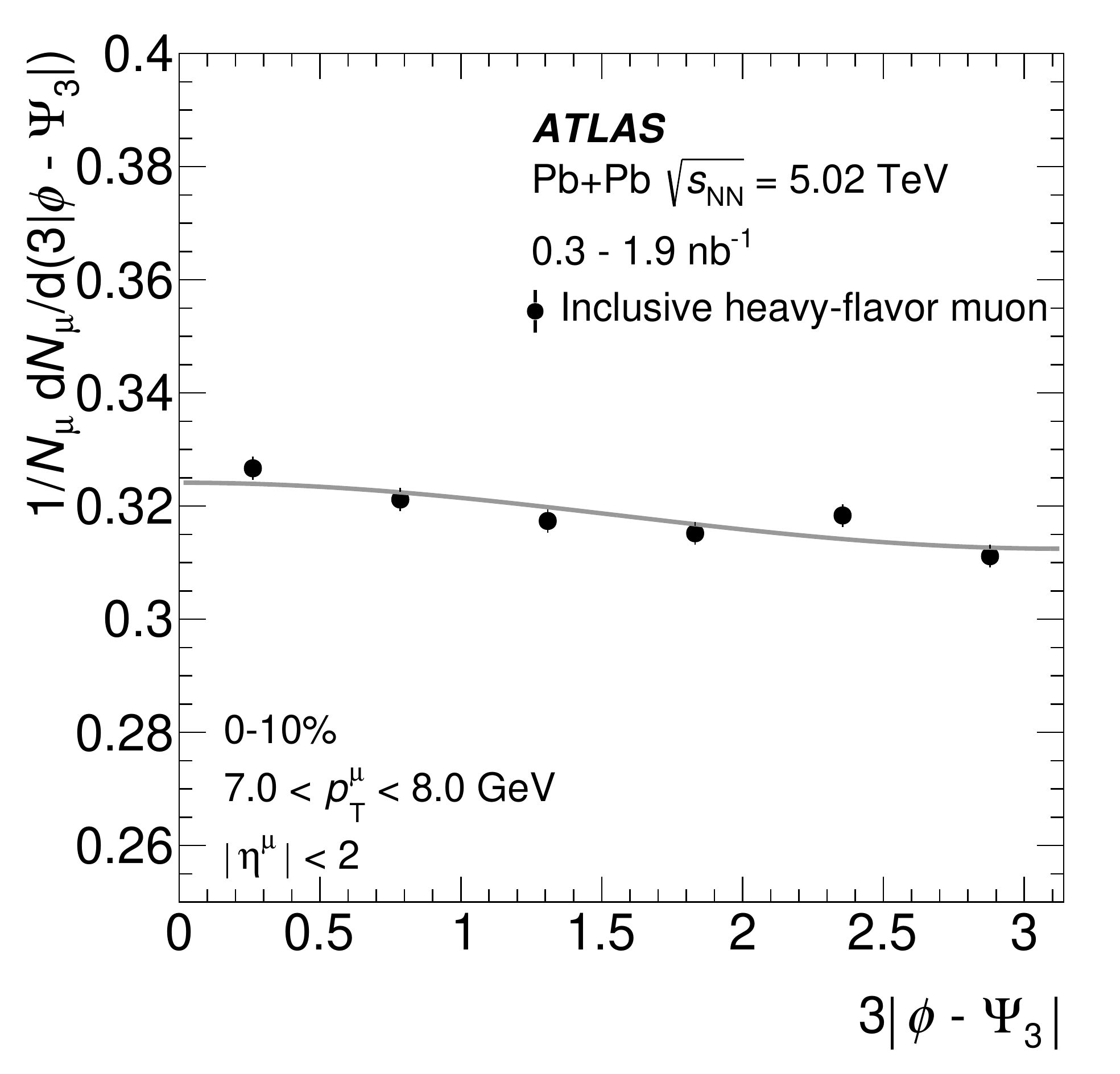}\\
\includegraphics[width=0.40\linewidth]{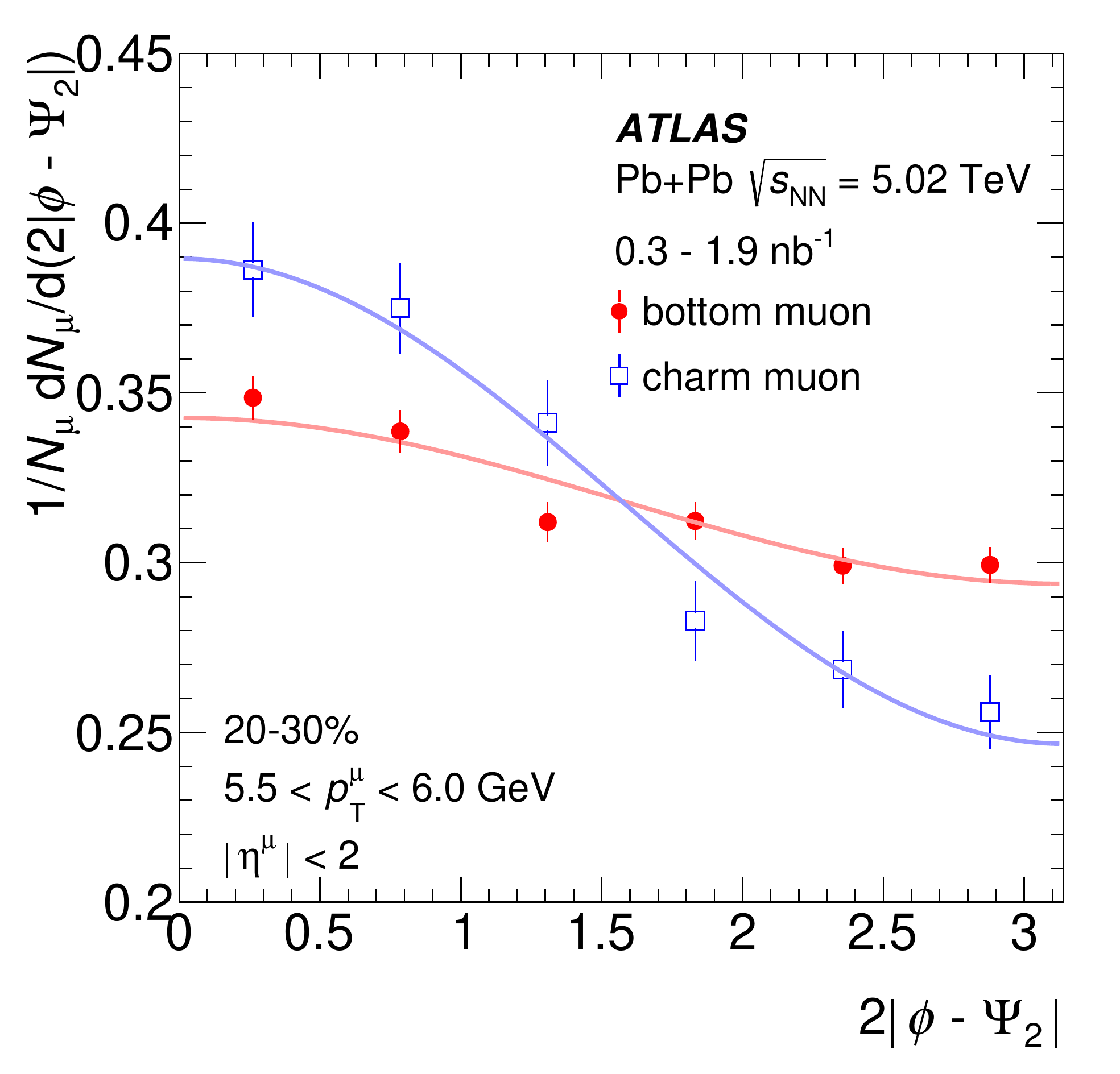}
\includegraphics[width=0.40\linewidth]{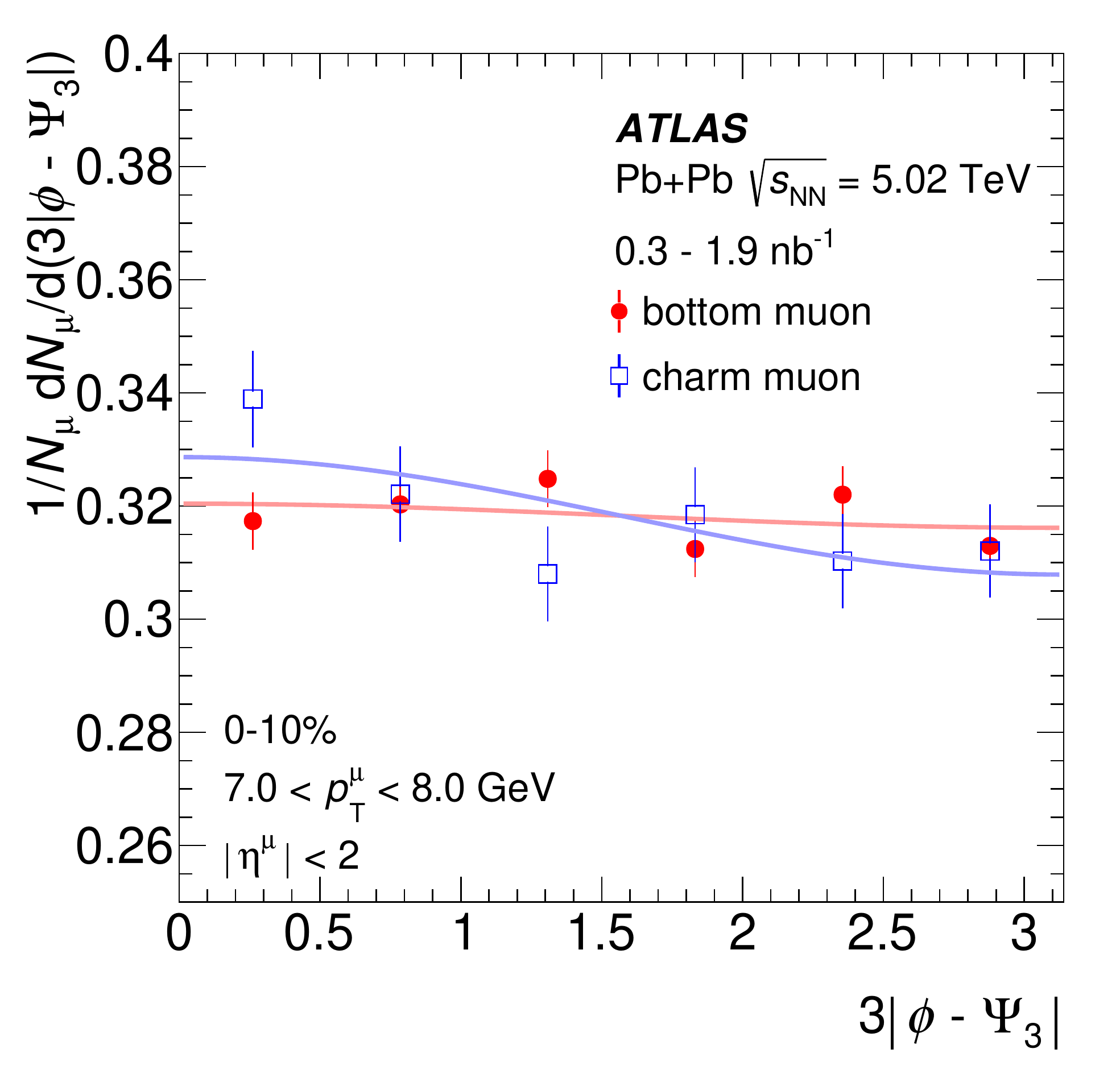}\\
\caption{
Examples of Fourier decomposition of inclusive heavy-flavour muon yields (top) and bottom/charm muon yields (bottom) in \PbPb collisions at $\sqn = 5.02~\TeV$ as a function of \dphiPsi{2} (left) and \dphiPsi{3} (right) for two selected intervals in muon \pt and centrality. The inclusive heavy-flavour muon yields are obtained from $\rho$ template fits,
while the bottom and charm muon yields are further separated using \dcaT fits.
In all cases the error bars on data indicate the statistical uncertainties obtained from the $\rho$ or \dcaT fit.
The lines indicate the extracted second (left) and third (right) Fourier harmonics.
}
\label{fig:exampleFlowFit_inclusive}
\end{figure}

The top row of Figure~\ref{fig:exampleFlowFit_inclusive} shows the inclusive heavy-flavour muon yield as a function of \dphiPsi{2} (left) and \dphiPsi{3} (right),
and the bottom row shows the charm and bottom muon yields as a function of \dphiPsi{2} (left) and \dphiPsi{3} (right).
The lines indicate the second (left) and third (right) extracted Fourier harmonics
from which the $v_2^\mathrm{raw}$ and
$v_3^\mathrm{raw}$ coefficients are extracted.

\section{Systematic uncertainties}
 
Systematic uncertainties are presented for different categories covering each step of the analysis procedure:
1) muon efficiency;
2) $\rho$ fit;
3) \dcaT fit;
4) light/onia background;
5) other background;
6) $\rho$--$\dcaT$ correlation;
7) event-plane resolution; and
8) jet bias.
Table~\ref{tab:systematics_summary} summarises the contributions from different sources of systematic uncertainty to the final flow-coefficient results.   Systematic uncertainties from all sources are summed in quadrature to determine the total uncertainty.

\begin{table}[htp]
\caption{\label{tab:systematics_summary} Summary of the typical sizes of the absolute systematic uncertainties of all categories for the flow coefficient results.
The \dcaT related systematic uncertainties are not relevant for the heavy-flavour inclusive flow measurements as the \dcaT fit is not utilised for these results.
Systematic uncertainties from the event-plane resolution and jet bias are negligible and not included in the final uncertainties, and therefore are not shown in the table.
The "$<$" symbol indicates the provided values are the maximum systematic variation for a given category.
}
\begin{center}
\renewcommand{\arraystretch}{1.4}
\resizebox{0.98\linewidth}{!}{
\begin{tabular}{l|c c c}
\hline
Category						&Inclusive heavy-flavour muon $v_2$ ($v_3$) 		&Charm muon $v_2$ ($v_3$)			& Bottom muon $v_2$ ($v_3$) \\
\hline\hline
Muon efficiency					&$<0.0002$ ($0.0006$)           &$<0.006$ ($0.001$)			    &$<0.001$ ($0.001$)\\
\hline
$\rho$ fit						&$<0.004$ ($0.006$)	            &$<0.009$ ($0.01$)~~		        &$<0.005$ ($0.003$)\\
\hline
$\dcaT$ fit						&N/A					        &$<0.02$  ($0.03$)			    &$<0.01$ ($0.01$)\\
\hline
Light/onia bkg		            &$<0.004$ ($0.002$)		        &$<0.02$  ($0.01$)			    &$<0.008$ ($0.004$)\\
\hline
Other bkg				        &$<0.000001$ ($0.000001$)		&$<0.002$ ($0.004$)		        &~~$<0.001$ ($0.0004$)\\
\hline
$\rho$--$\dcaT$ correlation		&N/A					        &~~$<0.01$  ($0.004$)		        &$<0.007$ ($0.005$)\\
\hline
\hline
\end{tabular}
}
\end{center}
\end{table}

The systematic uncertainties from the MS reconstruction efficiency and muon trigger efficiency corrections are dominated by the uncertainty in determining the data-to-MC scale factor. The scale factor uncertainties are evaluated following the procedure from previous ATLAS measurements~\cite{PERF-2015-10} including variations in the tag-and-probe efficiency extraction method, object-matching resolution, and purity of the sample.
The systematic uncertainty in the muon trigger efficiency also includes the determination of the centrality-dependent correction factors.
The uncertainty on the flow coefficients resulting from uncertainties in the muon ID reconstruction efficiency is evaluated by comparing the results with and without the ID efficiency correction, as the ID efficiency is approximately $99\%$ for all centralities.
The muon efficiency systematic uncertainties are correlated between the resulting charm and bottom muon results.
 
The systematic uncertainty of the $\rho$ fit is due to the uncertainty in the shifts and smearing parameters for single-muon momentum response in simulation, which is evaluated by comparing the nominal results with those 1) without any shifts or smearing, 2) only applying shifts and smearing to the signal muons, and
3) incorporating additional smearing of the background $\rho$ distributions in simulation to match data distributions in the background-enriched region ($\rho > 0.2$).
The changes resulting from these variations are combined in quadrature.
The combined uncertainty from the $\rho$ fit is $1\%$--$10\%$ for the charm and bottom muon results, depending on the muon \pt and $\eta$, but without dependence on centrality. The relative systematic variations are found to be the largest at low \pt and large $\eta$.
The $\rho$ fit systematic uncertainties are correlated between the resulting charm and bottom muon results.
 
The uncertainty in the \dcaT template shift and smearing parameters is tested by comparing results when determining the parameters using 2018 data (as in the nominal results) with results when using the 2015 \PbPb data to determine the parameters, which covers the slightly different detector alignment between the two data sets.
Sensitivity to the charm and bottom hadron \pt spectra reweighting in simulation is covered by a variation in which the \pt spectra are reweighted to agree with those from \PYTHIAV{8} simulations without any modification due to QGP.
The variations in the results due to \dcaT template shift and smearing and \pt spectra reweighting are considered to be uncorrelated and are combined in quadrature,
and the combined systematic uncertainty is $1\%$--$20\%$ for the charm and bottom muon results, depending on muon \pt and centrality.
The relative systematic variations are found to be the largest at high \pt and in more peripheral events.
The \dcaT fit systematic uncertainties are anti-correlated between the resulting charm and bottom muon results.
 
The magnitude of the light/onia contribution is held at a fixed fraction relative to the inclusive heavy-flavour muon contribution, based on the \PYTHIAV{8} MC sample.
The analysis is repeated to study this choice, first ignoring the light/onia contribution and then fixing it to twice the fraction predicted by \PYTHIAV{8}.
As is shown in Ref.~\cite{CMS-BPH-10-002}, \PYTHIA overestimates prompt quarkonium production at the LHC, and thus these variations of the light/onia contribution are large enough to not underestimate the uncertainty.
Each nominal result is assigned a systematic uncertainty equal to the larger of the changes from the two variations.
For the nominal results,  light/onia muons are assumed to have the same $v_2$ and $v_3$ as the inclusive heavy-flavour muons.
The analysis is repeated with variations assuming light/onia muons have zero flow coefficients or double the inclusive heavy-flavour muon flow coefficients.
The larger of the resulting changes is assigned as the systematic uncertainty.
The light/onia systematic uncertainties are anti-correlated between the resulting charm and bottom muon results.
 
The contributions of hadronic showers are ignored in the nominal analysis.
They were included in the analysis based on $\rho$ and \dcaT templates from the \PYTHIAV{8} simulation with a fixed relative fraction of a few percent also obtained from \PYTHIAV{8} simulation.
The deviation from the nominal result is found to be negligible for the inclusive heavy-flavour muon flow coefficients, and
approximately $8\%$ at low \pt ($< 8$~\GeV) and less than $1\%$ at high \pt ($> 12$~\GeV) for charm and bottom muon flow coefficients.
 
All muons are assumed to have independent $\rho$ and \dcaT distributions in the nominal results, which is only true for signal muons.
To test the sensitivity to this assumption,
\dcaT fits in data are repeated with a requirement of $\rho < 0.1$ on the sample, thus significantly reducing the background contribution.
The difference between results with and without the restriction on $\rho$ is assigned as a systematic uncertainty to cover the systematic effect of ignoring any correlation between $\rho$ and \dcaT for background muons.
 
The systematic uncertainty associated with the event-plane angle resolution is evaluated by measuring the event-plane resolution in two subregions of the FCal ($3.2 < |\eta| < 4.0$ and $4.0 < |\eta| < 4.8$), following a previous ATLAS flow analysis~\cite{HION-2016-06}.
The systematic uncertainties are evaluated independently for $\Psi_2$ and $\Psi_3$. The maximum difference between these two variations and the nominal results is considered as a systematic uncertainty.
The uncertainty associated with the event-plane angle resolution is found to be negligible compared to other systematic uncertainties, and thus is not included in the results.
 
The charm and bottom muons are often produced with a recoil jet.
The orientation of $\Psi_n$ could be biased to align with the signal muon if the recoil jet reaches the FCal~\cite{HION-2015-06}.
The magnitude of this bias in muon $v_2$ and $v_3$ is studied with \PYTHIA generator-level muon flow in samples overlaid with \PbPb data.
The bias is found to be 0.3\%--0.4\% for $v_2$ and $v_3$, and it is larger in peripheral events than in more-central events.  This small bias is negligible and is not included as a systematic uncertainty.

\section{Results}

\begin{figure}[htb]
\centering
\includegraphics[width=0.7\linewidth]{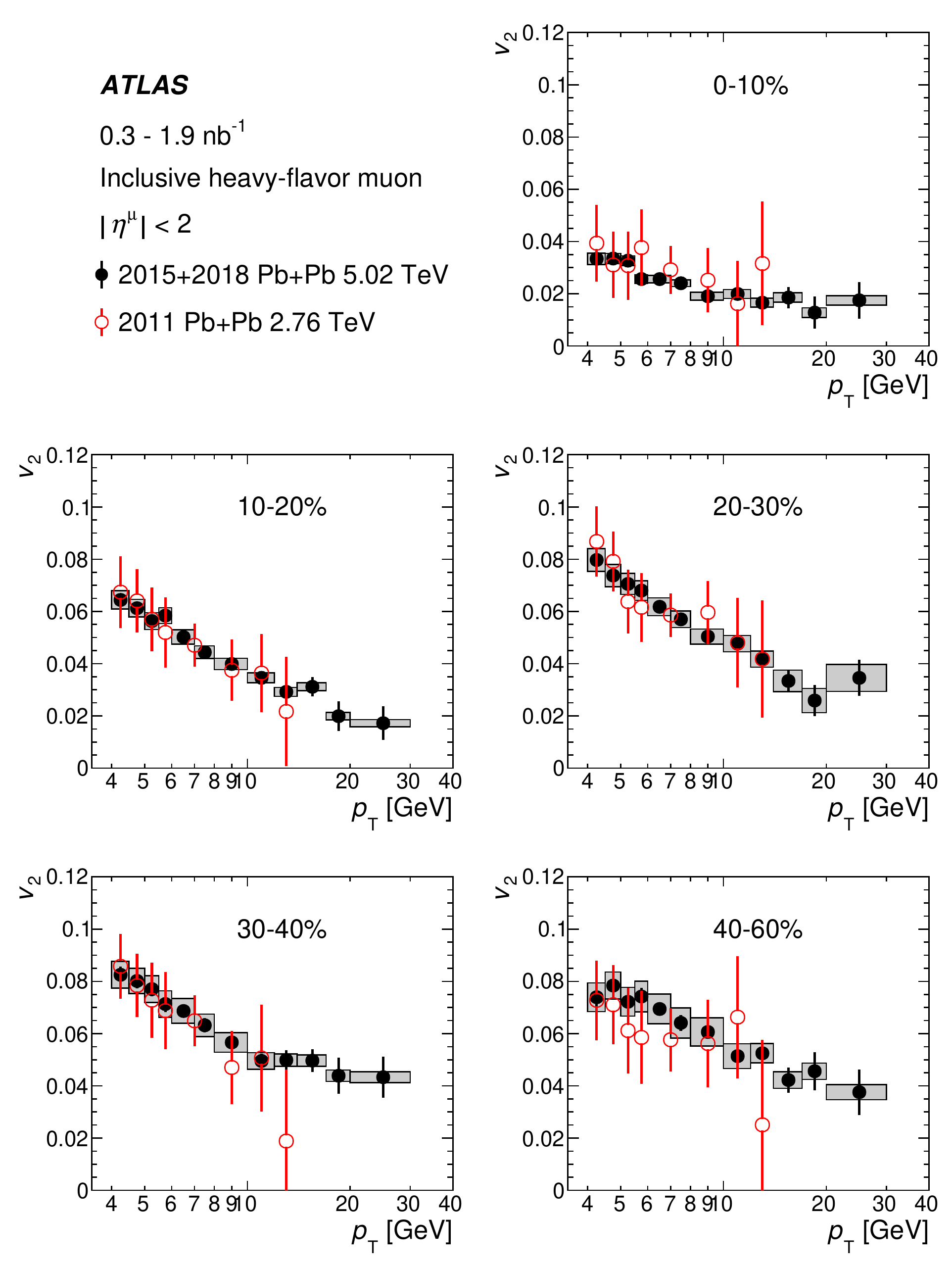}
 
\caption{Inclusive heavy-flavour muon $v_2$ as a function of \pt in the combined 2015 and 2018 $\sqn=5.02~\TeV$ \PbPb data compared with the results in the $\sqn=2.76~\TeV$ \PbPb data~\cite{HION-2015-06}.
Statistical uncertainties are shown as vertical lines and systematic uncertainties as boxes for the $\sqn=5.02~\TeV$ results.
For better visibility the statistical and systematic uncertainties of the $\sqn=2.76~\TeV$ data are combined in quadrature and shown as vertical lines.
Each panel represents a different centrality interval.}
\label{fig:v2_incl}
\end{figure}

\begin{figure}[htb]
\centering
\includegraphics[width=0.7\linewidth]{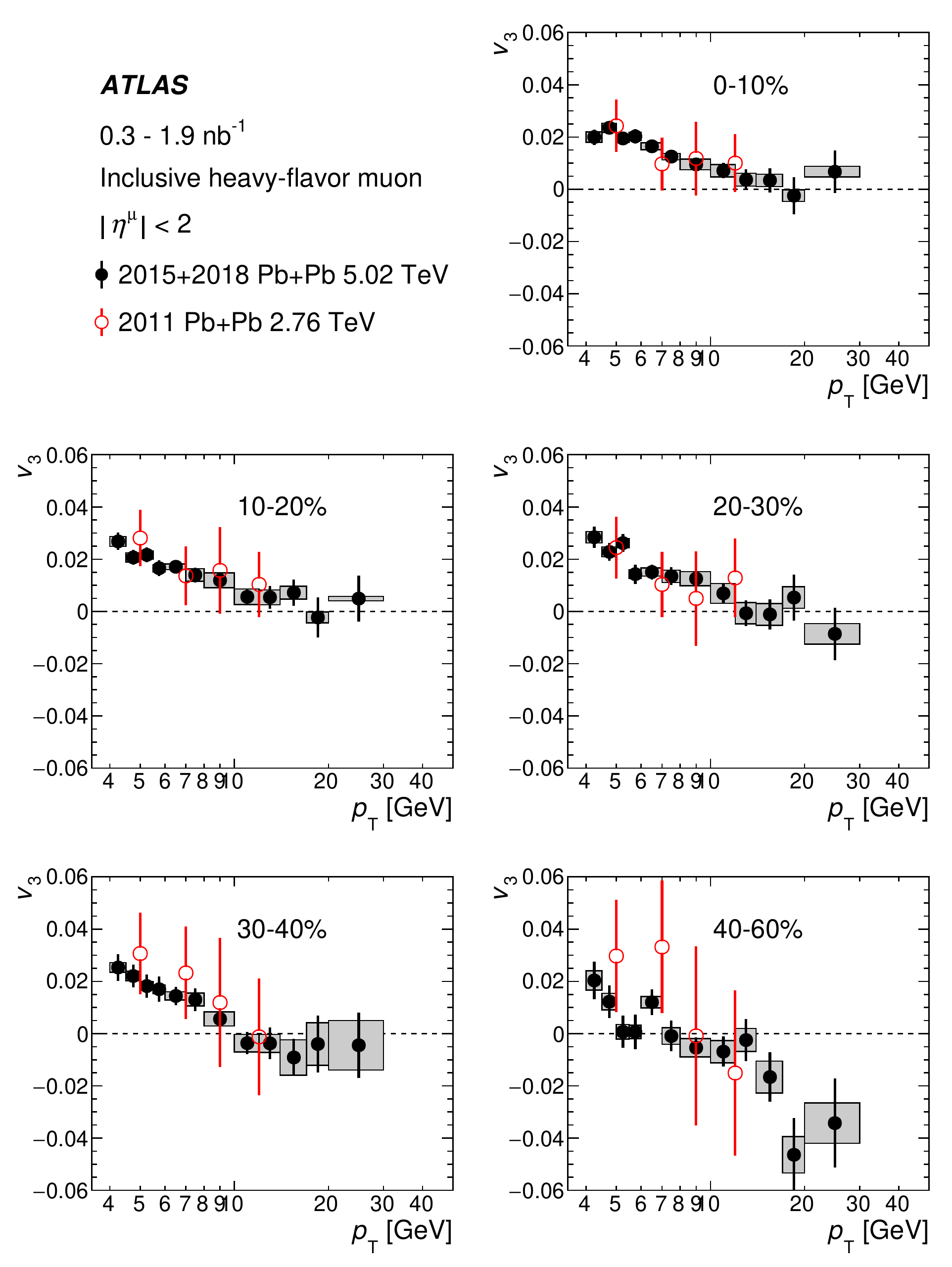}\\
\caption{Inclusive heavy-flavour muon $v_3$ as a function of \pt in the combined 2015 and 2018 $\sqn=5.02~\TeV$ \PbPb data compared with the results in the $\sqn=2.76~\TeV$ \PbPb data~\cite{HION-2015-06}.
Statistical uncertainties are shown as vertical lines and systematic uncertainties as boxes for the $\sqn=5.02~\TeV$ results.
For better visibility the statistical and systematic uncertainties of the $\sqn=2.76~\TeV$ data are combined in quadrature and shown as vertical lines.
Each panel represents a different centrality interval.}
\label{fig:v3_incl}
\end{figure}
 
Figure~\ref{fig:v2_incl} shows the inclusive heavy-flavour muon elliptic flow coefficient $v_2$ as a function of \pt in the $\sqn=5.02~\TeV$ \PbPb~ data.   Each panel corresponds to a different \PbPb~ centrality interval.   The
$v_2$ results decrease steadily with \pt over the entire \pT range and in all centrality intervals.   The
overall magnitude of $v_2$ is smaller in the most central 0--10\% selection, as expected since
the corresponding smaller impact parameter Pb+Pb collisions have smaller initial geometric ellipticity.
 
Figure~\ref{fig:v3_incl} shows the inclusive heavy-flavour muon triangular flow coefficient $v_3$ as a function of \pt in the $\sqn=5.02~\TeV$ \PbPb~ data.   Each panel corresponds to a different \PbPb~ centrality interval.   The $v_3$ results decrease steadily with \pt over the entire \pT range in all centrality intervals  within statistical and systematic uncertainties.   The
overall magnitude of $v_3$ is quite similar in all centrality intervals, as expected since triangular deformations of the initial geometry are primarily the result of fluctuations and are generally unrelated to any intrinsic geometry from the colliding nuclei~\cite{Alver:2010gr}.
 
Each panel of \Figs{\ref{fig:v2_incl}}{\ref{fig:v3_incl}} also presents previous ATLAS results from $\sqn=2.76~\TeV$ \PbPb data~\cite{HION-2015-06}.
Compared to the earlier result, the present $\sqn=5.02~\TeV$ results significantly improve the statistical precision of the measurements and extend the \pt~range.
The $\sqn=5.02~\TeV$ and $\sqn=2.76~\TeV$ \PbPb data inclusive heavy-flavour muon $v_2$ and $v_3$ coefficients are consistent with each other within uncertainties.
Indeed, according to hydrodynamical models, no significant differences are expected between \PbPb~collisions at the two different energies~\cite{Zhao2017}.  Thus the observed consistency between $\sqn=2.76~\TeV$ and $\sqn=5.02~\TeV$ data is in agreement with expectations.   The
inclusive charged-particle flow coefficients are also observed to be nearly identical at the two collision energies~\cite{HION-2016-06}.

\begin{figure}[htb]
\centering
\includegraphics[width=0.7\linewidth]{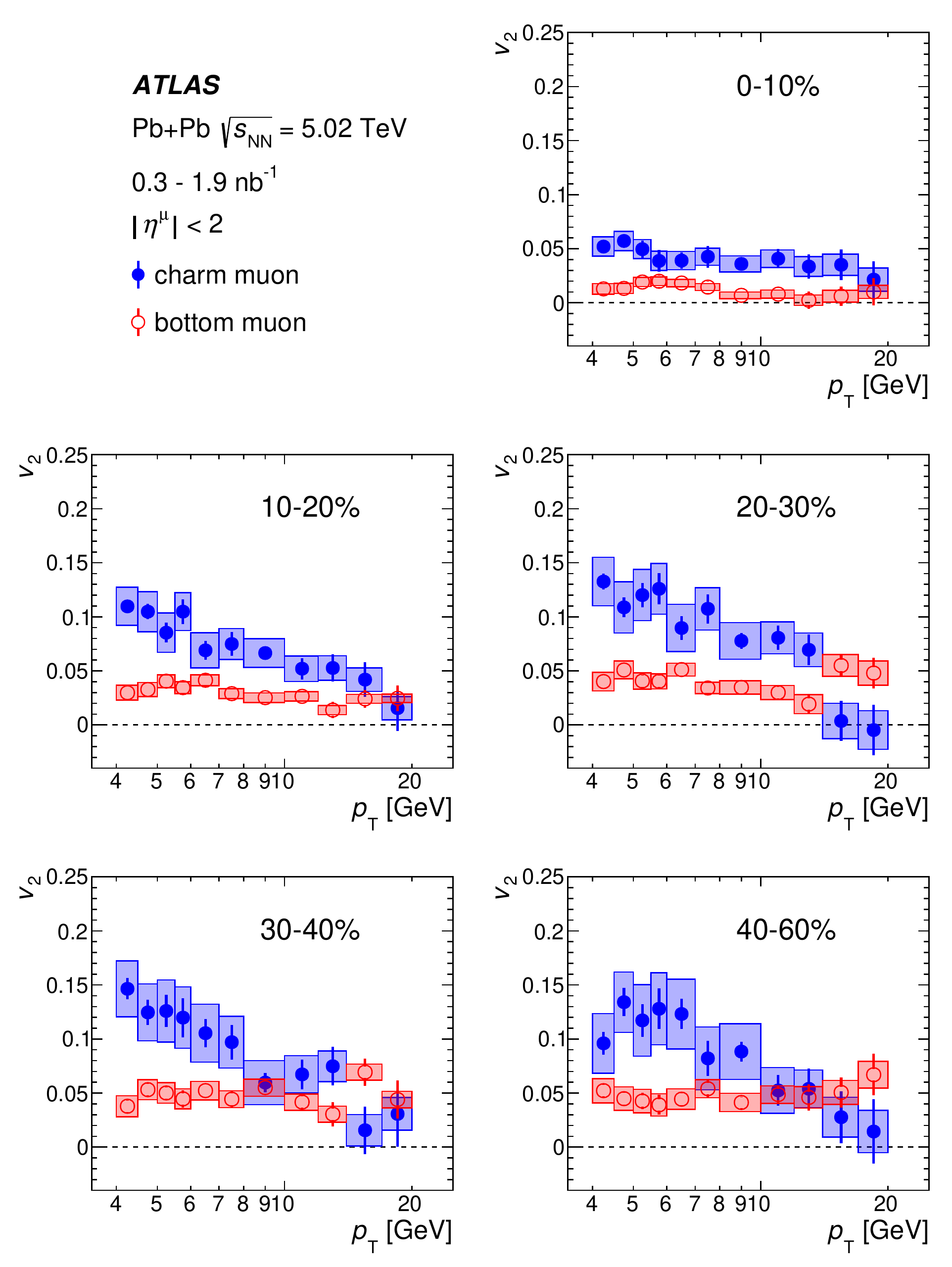}
\caption{Charm and bottom muon $v_2$ as a function of \pt in the combined 2015 and 2018 $\sqn=5.02~\TeV$ \PbPb data.
Statistical uncertainties are shown as vertical lines and systematic uncertainties as boxes for charm and bottom muons.   The charm and bottom uncertainties are partially anti-correlated.
Each panel presents a different centrality interval.}
\label{fig:v2_cb}
\end{figure}

Figure~\ref{fig:v2_cb} shows the separated charm and bottom muon $v_2$ as a function of \pt, with each panel presenting a different \PbPb~centrality interval.
The charm and bottom flow coefficients are extracted only up to $\pt = 20$~\GeV, since above that transverse momentum range the inclusive heavy-flavour $v_2$ values are small and the charm-to-bottom separation procedure becomes sensitive to fluctuations in data and yields unstable results.
The results indicate a non-zero $v_2$ for both the charm and bottom muons, with substantially larger elliptic flow coefficients for charm muons.    The statistical and systematic uncertainties have a significant contribution that is anti-correlated between the charm and bottom $v_{2}$, i.e.\ an upward fluctuation in the charm $v_{2}$ in a particular \pt bin will be correlated with a downward fluctuation in the bottom $v_{2}$ in the same bin and vice versa.
For $\pt < 14~\GeV$, both charm and bottom muon $v_2$ increase from central to mid-central events, reaching
maximum between $20\%$ and $40\%$ centrality. Over the range of $\pt > 14~\GeV$, the charm and bottom muon $v_2$ show no obvious centrality dependence with larger uncertainties.

Qualitatively, the charm and bottom $v_{2}$ ordering  matches theoretical expectations where the heavier bottom quarks have a smaller modification to their initial momentum trajectories due to their larger masses.
Light quarks and heavy quarks can lose energy in traversing the QGP via induced gluon
radiation~\cite{Baier:1994bd};
however, heavy quarks with momentum
less than or approximately equal to
the quark mass ($p \lesssim m$)
radiate less than light quarks due to a suppression of radiation at small angles relative to the quark direction, referred to as the `dead-cone' effect~\cite{Dokshitzer:2001zm}.
Thus, at high $\pT >> m$, all quark flavors should lose comparable energy in the QGP, while at lower \pT there should be a hierarchy with light quarks losing the most energy, then charm quarks, and finally bottom quarks losing the least energy.
Thus the heavier bottom quarks with $\pT \lesssim 20-30$~GeV are expected to lose less energy in the QGP and thus have a smaller azimuthal anisotropy.
 
\begin{figure}[htb]
\centering
\includegraphics[width=0.7\linewidth]{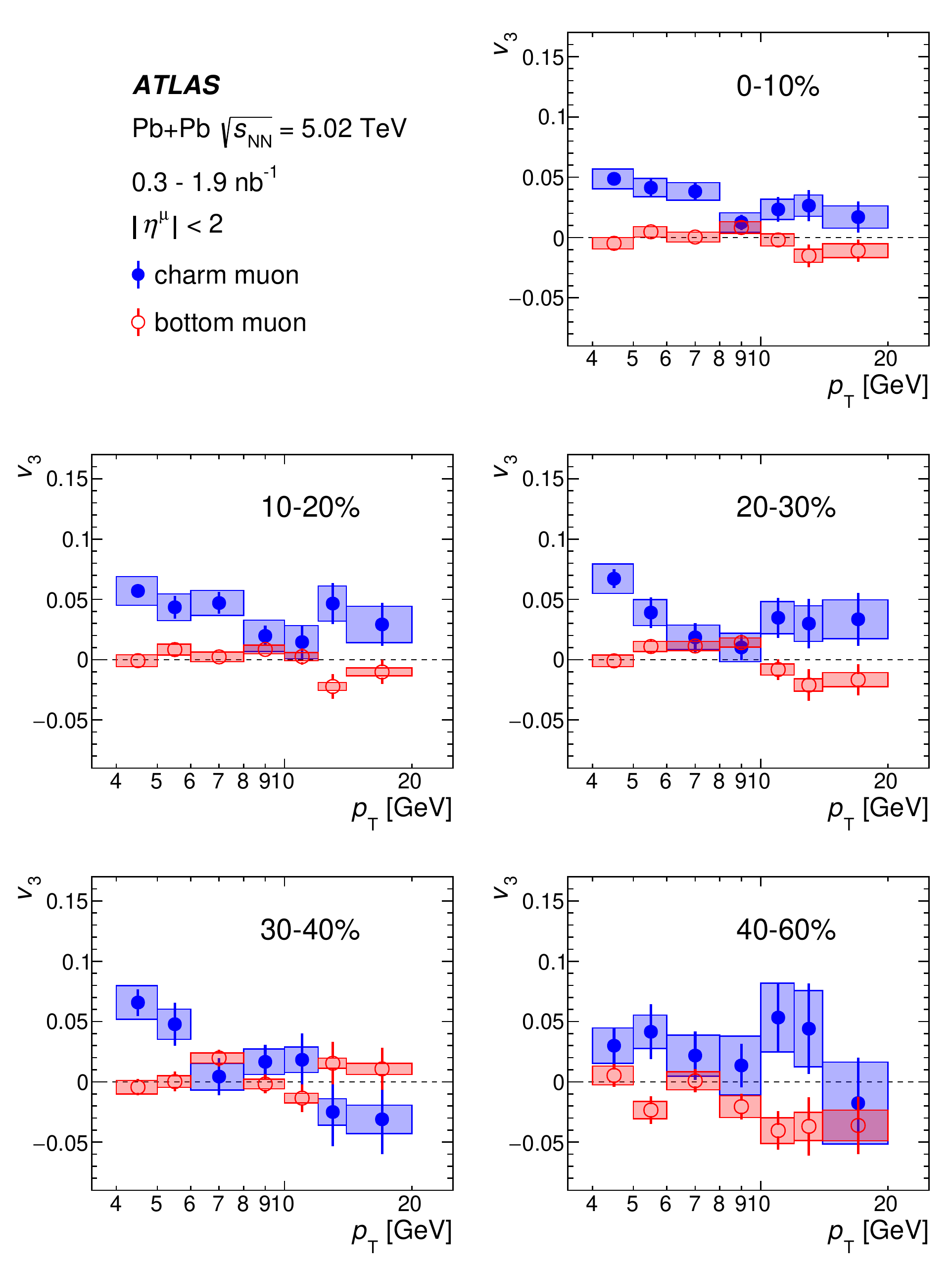}
\caption{Charm and bottom muon $v_3$ as a function of \pt in the combined 2015 and 2018 $\sqn=5.02~\TeV$ \PbPb data.
Statistical uncertainties are shown as vertical lines and systematic uncertainties as boxes for charm and bottom muons.
The charm and bottom uncertainties are partially anti-correlated.
Each panel presents a different centrality interval.}
\label{fig:v3_cb}
\end{figure}

Figure~\ref{fig:v3_cb} shows the separated charm and bottom muon $v_3$ as a function of \pt, with each panel presenting a different \PbPb~centrality interval.
The charm muons show significant non-zero $v_3$ values, which are independent of centrality.
The bottom muons have $v_3$ values that are significantly below that of charm muons at all \pT and in all centrality intervals.

\begin{figure}
\centering
\includegraphics[width=0.49\linewidth]{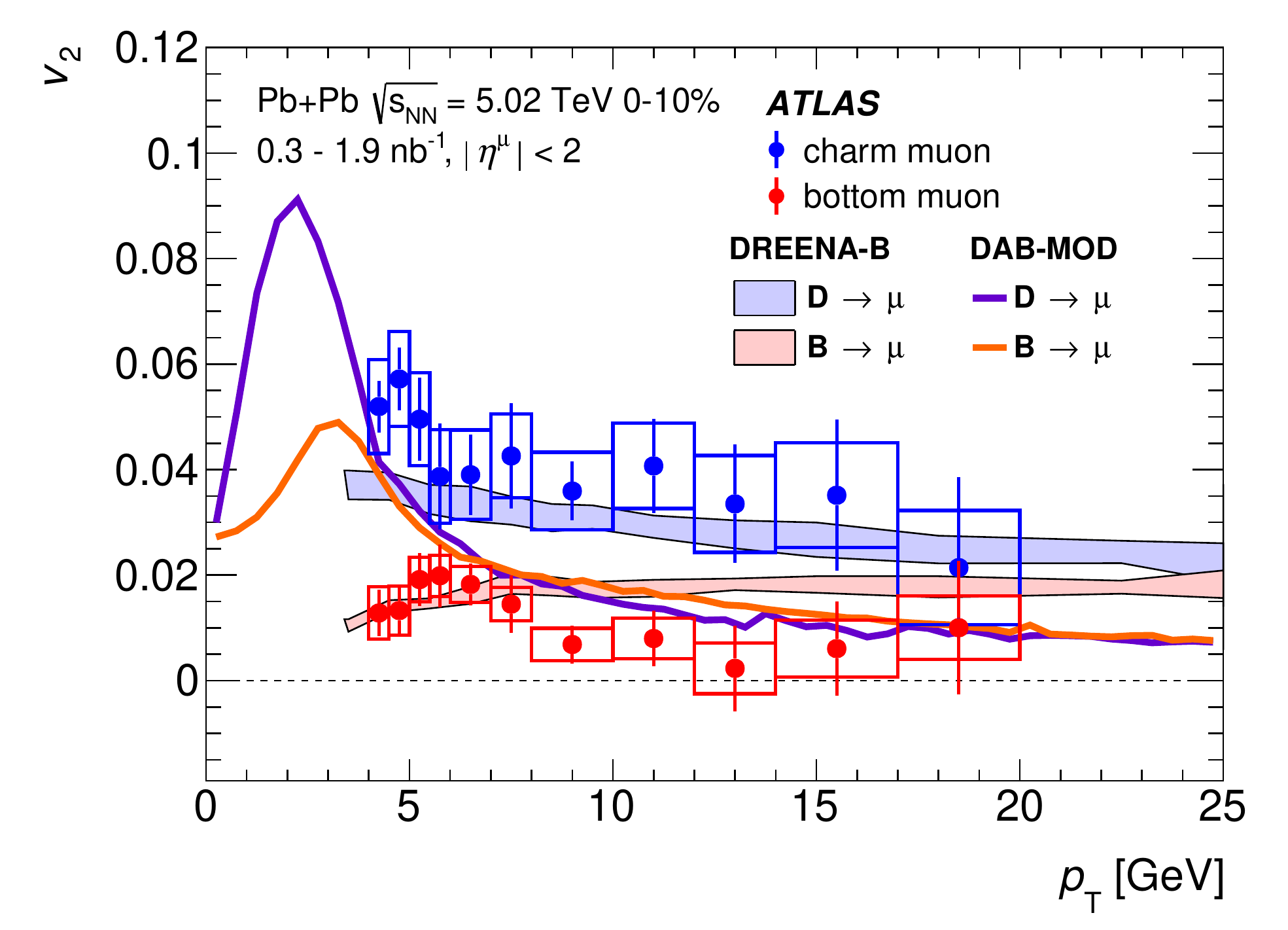}
\includegraphics[width=0.49\linewidth]{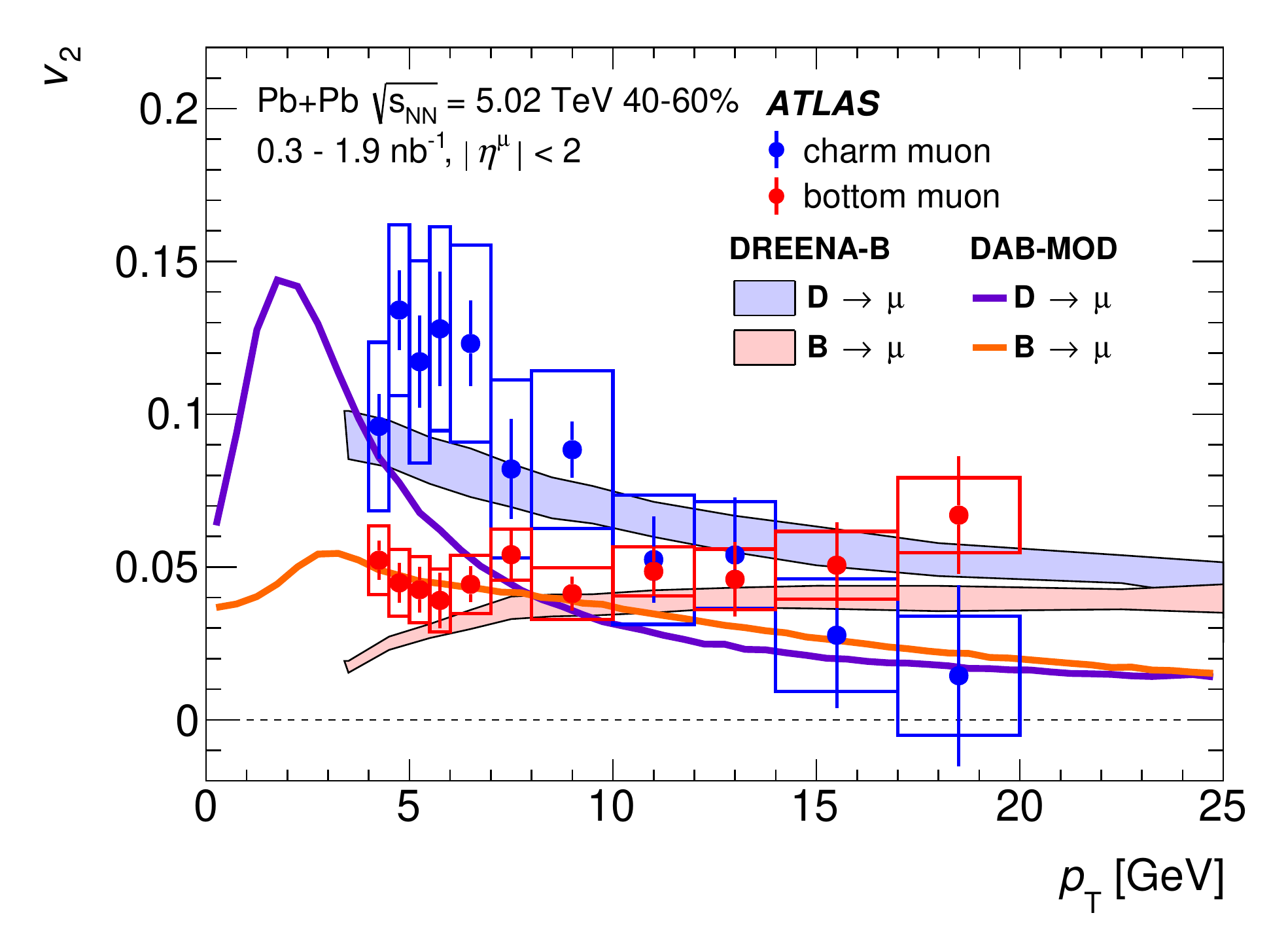}
\caption{Charm and bottom muon $v_2$ as a function of \pt in the $\sqn=5.02~\TeV$ \PbPb data for the 0--10\% (left) and 40--60\% (right) centrality interval, compared with theoretical predictions based on \textsc{dreena-b}~\cite{Zigic:2018ovr} and \textsc{dab-mod}~\cite{Katz:2019fkc,Prado:2016szr} in the same centrality intervals for charm and bottom muon $v_2$.
For the data, statistical uncertainties are shown as vertical lines and systematic uncertainties as boxes.
The charm and bottom uncertainties are partially anti-correlated.
}
\label{fig:theory1}
\end{figure}

Figure~\ref{fig:theory1} shows the results for $v_2$ versus \pt, from Figure~\ref{fig:v2_cb}, compared with theoretical calculations: \textsc{dreena-b} from Ref.~\cite{Zigic:2018ovr}, and \textsc{dab-mod} from Refs.~\cite{Katz:2019fkc,Prado:2016szr} for charm and bottom decay muons in the Pb+Pb 0--10\% (left) and 40--60\% (right) centrality intervals.
The \textsc{dreena-b} calculation includes radiative and collisional energy loss of the heavy quarks traversing the QGP, the latter modelled via a 1~+~1D Bjorken expansion~\cite{PhysRevD.27.140}.
The \textsc{dreena-b} theoretical uncertainties reflect the range of magnetic to electric screening masses as constrained by non-perturbative calculations~\cite{PhysRevD.27.140}.
The predicted $D$ meson $v_2$ is higher than the $B$ meson $v_2$, with the two converging at
$\pt \approx 25~\GeV$ as expected when the \pt is much larger than the charm and bottom quark masses.   Using \PYTHIAV{8} for decay kinematics, the charm muon and bottom muon $v_2$ results are calculated and shown.
The predominant effect in going from the parent meson $v_2$(\pt) to the daughter muon $v_2$(\pt) is a shift downward in \pt.   The predictions are in reasonable agreement with the experimental data, although they overestimate the $v_2$ at high \pt of bottom muons in central events.
The \textsc{dab-mod} framework includes calculations with only Langevin drag and diffusion contributions.
The curves shown here are obtained with \textsc{Trento} geometric initial conditions~\cite{PhysRevC.92.011901}, heavy-quark Langevin dynamics with the Moore and Teaney parameterisation~\cite{Moore:2004tg}, and coupling values for charm (bottom) of $D/2\pi T$ = 2.23 (2.79), where $D$ is the spatial diffusion coefficient and $T$ is the temperature.   The
decoupling temperature of heavy quarks from the medium
is $T = 160~\mathrm{MeV}$ and both coalescence and fragmentation are implemented for hadronisation.
The \textsc{dab-mod} predictions with only Langevin dynamics are roughly a factor of three (two) lower for charm (bottom) muons compared with \textsc{dreena-b}.
Additional energy loss contributions to \textsc{dab-mod}, not included here, tend to increase the high $\pt$ anisotropies.
At lower $\pt$, the \textsc{dreena-b} $v_2$ results rise significantly.   A key component of these calculations is the modelling of the QGP transverse expansion, and thus it will be instructive in the future to compare the calculations with a common QGP model to test whether the differences arise from the QGP modeling or the energy-loss implementation.

\FloatBarrier
\section{Conclusion}

In summary, a measurement of elliptic and triangular flow coefficients for heavy-flavour decay muons in Pb+Pb collisions at
$\sqn=5.02~\TeV$ is presented, including a separation between charm and bottom contributions. The measurement
uses a combined 2015 and 2018 data set corresponding to a total integrated luminosity of up to $1.9~\mathrm{nb}^{-1}$ recorded by the ATLAS experiment at the LHC.
The inclusive heavy-flavour muon $v_2$ and $v_3$ values measured in $4<\pt<30~\GeV$ are observed to decrease with \pt for all centrality intervals.
The $v_2$ and $v_3$ values are consistent within uncertainties with previous Pb+Pb measurements at $\sqn=2.76~\TeV$.
Further separating the inclusive heavy-flavour muons into charm and bottom muons reveals a significantly larger $v_{2}$ and $v_{3}$ for charm muons than for bottom muons.    The results indicate that while both the charm and bottom quarks have their trajectories and momenta modified when traversing the QGP, the effect is stronger for the
charm quarks.
At fixed \pt, the charm and bottom muon $v_2$ values decrease from 40-60\% to 0-10\% centrality intervals as observed in charged hadron elliptic-flow measurements~\cite{HION-2011-01}.
Theoretical calculations have a similar qualitative trend with smaller flow coefficients for  muons from decays of the heavier bottom quarks.   The results can significantly discriminate between models of heavy-quark energy loss and constrain heavy-quark transport coefficients in the QGP.

\section*{Acknowledgements}

% The next lines are included from the .//acknowledgements/Acknowledgements.tex input file

We thank CERN for the very successful operation of the LHC, as well as the
support staff from our institutions without whom ATLAS could not be
operated efficiently.
 
We acknowledge the support of ANPCyT, Argentina; YerPhI, Armenia; ARC, Australia; BMWFW and FWF, Austria; ANAS, Azerbaijan; SSTC, Belarus; CNPq and FAPESP, Brazil; NSERC, NRC and CFI, Canada; CERN; CONICYT, Chile; CAS, MOST and NSFC, China; COLCIENCIAS, Colombia; MSMT CR, MPO CR and VSC CR, Czech Republic; DNRF and DNSRC, Denmark; IN2P3-CNRS and CEA-DRF/IRFU, France; SRNSFG, Georgia; BMBF, HGF and MPG, Germany; GSRT, Greece; RGC and Hong Kong SAR, China; ISF and Benoziyo Center, Israel; INFN, Italy; MEXT and JSPS, Japan; CNRST, Morocco; NWO, Netherlands; RCN, Norway; MNiSW and NCN, Poland; FCT, Portugal; MNE/IFA, Romania; MES of Russia and NRC KI, Russia Federation; JINR; MESTD, Serbia; MSSR, Slovakia; ARRS and MIZ\v{S}, Slovenia; DST/NRF, South Africa; MINECO, Spain; SRC and Wallenberg Foundation, Sweden; SERI, SNSF and Cantons of Bern and Geneva, Switzerland; MOST, Taiwan; TAEK, Turkey; STFC, United Kingdom; DOE and NSF, United States of America. In addition, individual groups and members have received support from BCKDF, CANARIE, Compute Canada and CRC, Canada; ERC, ERDF, Horizon 2020, Marie Sk{\l}odowska-Curie Actions and COST, European Union; Investissements d'Avenir Labex, Investissements d'Avenir Idex and ANR, France; DFG and AvH Foundation, Germany; Herakleitos, Thales and Aristeia programmes co-financed by EU-ESF and the Greek NSRF, Greece; BSF-NSF and GIF, Israel; CERCA Programme Generalitat de Catalunya and PROMETEO Programme Generalitat Valenciana, Spain; G\"{o}ran Gustafssons Stiftelse, Sweden; The Royal Society and Leverhulme Trust, United Kingdom.
 
The crucial computing support from all WLCG partners is acknowledged gratefully, in particular from CERN, the ATLAS Tier-1 facilities at TRIUMF (Canada), NDGF (Denmark, Norway, Sweden), CC-IN2P3 (France), KIT/GridKA (Germany), INFN-CNAF (Italy), NL-T1 (Netherlands), PIC (Spain), ASGC (Taiwan), RAL (UK) and BNL (USA), the Tier-2 facilities worldwide and large non-WLCG resource providers. Major contributors of computing resources are listed in Ref.~\cite{ATL-SOFT-PUB-2020-001}.
 
% End of text imported from the .//acknowledgements/Acknowledgements.tex input file

\printbibliography
 
\clearpage
% ATLAS Collaboration author list
% Reference date of HION-2019-11 is 2019-10-16
% Author list last updated on date 09-JUN-20
% Data extracted on 09-Jun-2020 for paper reference HION-2019-11
% at 10:22am
 
\begin{flushleft}
\hypersetup{urlcolor=black}
{\Large The ATLAS Collaboration}

\bigskip

\AtlasOrcid[0000-0002-6665-4934]{G.~Aad}$^\textrm{\scriptsize 102}$,    
\AtlasOrcid[0000-0002-5888-2734]{B.~Abbott}$^\textrm{\scriptsize 128}$,    
\AtlasOrcid{D.C.~Abbott}$^\textrm{\scriptsize 103}$,    
\AtlasOrcid[0000-0002-2788-3822]{A.~Abed~Abud}$^\textrm{\scriptsize 36}$,    
\AtlasOrcid[0000-0002-1002-1652]{K.~Abeling}$^\textrm{\scriptsize 53}$,    
\AtlasOrcid[0000-0002-2987-4006]{D.K.~Abhayasinghe}$^\textrm{\scriptsize 94}$,    
\AtlasOrcid[0000-0002-8496-9294]{S.H.~Abidi}$^\textrm{\scriptsize 166}$,    
\AtlasOrcid[0000-0002-8279-9324]{O.S.~AbouZeid}$^\textrm{\scriptsize 40}$,    
\AtlasOrcid{N.L.~Abraham}$^\textrm{\scriptsize 155}$,    
\AtlasOrcid[0000-0001-5329-6640]{H.~Abramowicz}$^\textrm{\scriptsize 160}$
% ...
\bigskip
\\

% ...
$^{36}$CERN, Geneva; Switzerland.\\
% ...
$^{40}$Niels Bohr Institute, University of Copenhagen, Copenhagen; Denmark.\\
% ...
$^{53}$II. Physikalisches Institut, Georg-August-Universit\"{a}t G\"ottingen, G\"ottingen; Germany.\\
% ...
$^{94}$Department of Physics, Royal Holloway University of London, Egham; United Kingdom.\\
% ...
$^{102}$CPPM, Aix-Marseille Universit\'e, CNRS/IN2P3, Marseille; France.\\
$^{103}$Department of Physics, University of Massachusetts, Amherst MA; United States of America.\\
% ...
$^{128}$Homer L. Dodge Department of Physics and Astronomy, University of Oklahoma, Norman OK; United States of America.\\
% ...
$^{155}$Department of Physics and Astronomy, University of Sussex, Brighton; United Kingdom.\\
% ...
$^{160}$Raymond and Beverly Sackler School of Physics and Astronomy, Tel Aviv University, Tel Aviv; Israel.\\
% ...
$^{166}$Department of Physics, University of Toronto, Toronto ON; Canada.\\
% ...

\end{flushleft}

% Created with Glance <Atlas.Glance@cern.ch>

\end{document}